\documentclass[aps,prx,reprint, superscriptaddress]{revtex4-2}
\usepackage{blindtext}
\usepackage{graphicx}
\usepackage{xcolor}
\usepackage{dcolumn}
\usepackage{bm}
\usepackage[utf8]{inputenc}

\begin{document}
	
	\preprint{APS/123-QED}
	
	\title{Emergent surface resonance from charge density wave symmetry breaking in TiSe$_2$}

\author{Turgut Yilmaz}
	\email{trgt2112@gmail.com}
	\affiliation{Department of Physics, Xiamen University Malaysia, Sepang 43900, Malaysia}
	\affiliation{Department of Physics, University of Connecticut, Storrs, CT 06269, USA}

\author{Yi Sheng Ng}
	\affiliation{Department of New Energy Science and Engineering, Xiamen University Malaysia, Sepang 43900, Malaysia}

\author{Muhammad Awais Fiaz}
	\affiliation{Department of Physics and Astronomy, University of New Hampshire, Durham, NH 03824, USA}

\author{Anil Rajapitamahuni}
    \affiliation{Department of Physics, SRM University - AP, Amaravati, Andhra Pradesh, 522502,India}
    \affiliation{Department of Physics and Astronomy, University of Nebraska-Lincoln, Nebraska 68588, USA}

\author{Asish K. Kundu}
	\affiliation{National Synchrotron Light Source II, Brookhaven National Lab, Upton, New York 11973, USA}

\author{Shawna M. Hollen}
	\affiliation{Department of Physics and Astronomy, University of New Hampshire, Durham, NH 03824, USA}

\author{Polina M. Sheverdyaeva}
	\affiliation{Consiglio Nazionale delle Ricerche, Istituto di Struttura della Materia (CNR-ISM), SS 14, km 163,5 I-34149 Trieste, Italy}

\author{Paolo Moras}
	\affiliation{Consiglio Nazionale delle Ricerche, Istituto di Struttura della Materia (CNR-ISM), SS 14, km 163,5 I-34149 Trieste, Italy}

\author{Ivana Vobornik}
	\affiliation{CNR-Istituto Officina dei Materiali (CNR-IOM), SS 14, km 163,5 I-34149 Trieste, Italy}

\author{Jun Fujii}
	\affiliation{CNR-Istituto Officina dei Materiali (CNR-IOM), SS 14, km 163,5 I-34149 Trieste, Italy}

\author{Shinichiro Ideta}
	\affiliation{Graduate School of Advanced Science and Engineering, Hiroshima University, Higashi-Hiroshima 739-8526, Japan}
	\affiliation{Research Institute for Synchrotron Radiation Science (HiSOR), Hiroshima University, Higashi-Hiroshima 739-0046, Japan}

\author{Kenya Shimada}
	\affiliation{Graduate School of Advanced Science and Engineering, Hiroshima University, Higashi-Hiroshima 739-8526, Japan}
	\affiliation{Research Institute for Synchrotron Radiation Science (HiSOR), Hiroshima University, Higashi-Hiroshima 739-0046, Japan}
	\affiliation{Research Institute for Semiconductor Engineering (RISE), Hiroshima University, 2-313 Kagamiyama, Higashi-Hiroshima 739-8527, Japan}
	\affiliation{The International Institute for Sustainability with Knotted Chiral Meta Matter (WPI-SKCM2), Hiroshima University, Higashi-Hiroshima 739-8526, Japan}

\author{Boris Sinkovic}
	\affiliation{Department of Physics, University of Connecticut, Storrs, CT 06269, USA}

\author{Elio Vescovo}
	\affiliation{National Synchrotron Light Source II, Brookhaven National Lab, Upton, New York 11973, USA}
	
\author{Hui-Qiong Wang}
	\affiliation{Department of Physics, Xiamen University Malaysia, Sepang 43900, Malaysia}

\author{Jin-Cheng Zheng}
\email{jczheng@xmu.edu.my}
	\affiliation{Department of Physics, Xiamen University Malaysia, Sepang 43900, Malaysia}

\date{\today}

\begin{abstract}
	Surface confined electronic states provide a fertile ground for discovering emergent phenomena that have no counterpart in the bulk, offering new routes to manipulate correlations, symmetry breaking, and dimensionality at the atomic scale. Here, we show that charge density wave (CDW) symmetry breaking can yield a surface states in 1T-TiSe$_2$. Micro angle resolved photoemission spectroscopy ($\mu$-ARPES) resolves a sharp, two dimensional surface resonant state (SRS) that emerges within the CDW reconstructed low energy spectrum. The SRS exhibits notable temperature dependence and its spectral weight collapses around $\sim$160 K, while CDW transition temperature $T_{CDW}$ is commonly reported as $\approx 202$~K. Slab DFT+$U$ calculations reproduce a surface localized resonance when CDW folding brings valence and conduction states into near degeneracy, suggesting a correlation tuned, surface selective origin. These results point to a form of correlation-tuned surface resonance in a layered CDW compound and suggest a framework for engineering low dimensional quantum states in van der Waals materials via symmetry breaking and electronic structure tuning.
\end{abstract}

\maketitle

\section{Introduction}

The surfaces of correlated quantum materials can host emergent electronic states that have no analogue in the bulk, often stabilized by reduced dimensionality, modified electronic environment, or reconstruction of the near surface atomic lattice. These surface specific states have been widely discussed in topological materials, Mott systems, and superconductors~\cite{damascelli2003angle,hasan2010colloquium,ando2013topological}, yet far less is known about analogous states arising from the interplay between CDW and electronic correlations. CDWs, common in layered transition metal dichalcogenides (TMDCs) and rare earth tritellurides, combine a periodic lattice distortion (PLD) with Fermi surface reconstruction, producing backfolded bands and gapped spectra~\cite{rossnagel2011origin,monceau2012electronic}. While this reconstruction is usually considered in terms of bulk properties, it naturally alters the near surface potential, enabling surface confined states through modified hybridization pathways.

Among these materials, 1T-TiSe$_2$ stands out as a prototypical system for exploring CDW physics, owing to its simple stoichiometry and a relatively high transition temperature of $T_{CDW} \approx 202$~K~\cite{di1976electronic} in addition to a long-standing transport anomaly near 160~K~\cite{rossnagel2002charge,miyahara1996tunnelling,craven1978mechanisms} that has remained elusive so far. Despite decades of experimental and theoretical efforts, the origin of the CDW remains controversial, with proposed mechanisms including excitonic condensation, electron-phonon interaction, and Jahn–Teller–like distortions~\cite{cercellier2007evidence,monney2009temperature,van2010exciton,hughes1977structural}. While the role of electronic interactions in 1T-TiSe$_2$ remains contested, it motivates closer scrutiny of low energy band structure tuning and symmetry breaking, thereby positioning 1T-TiSe$_2$ as a useful prototype for surface electronic structure in the presence of broken symmetry and folded bands.

\begin{figure*}[t]
	\centering
	\includegraphics[width=0.75\textwidth]{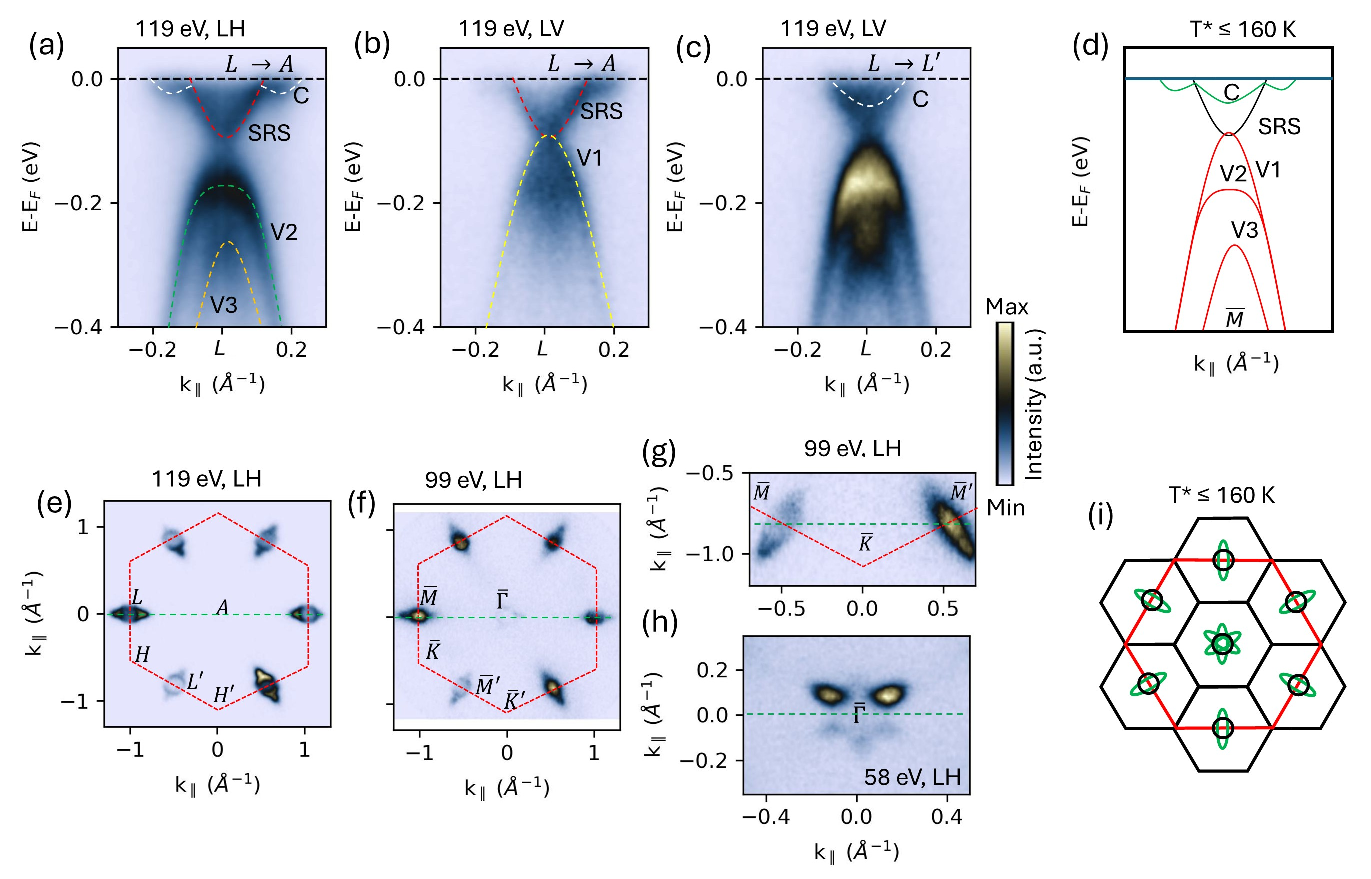} 
	
	\caption{(a) ARPES intensity plot measured along the $L$–$A$. The SRS appears as a sharp V-shaped dispersion (red dashed), coexisting with the bulk conduction band C (white dashed) and folded valence bands (V2, V3). (b) Same as (a) but measured with LV polarization. The SRS is selectively enhanced, while the folded valence bands (V1) are more clearly resolved. (c) ARPES intensity plot measured along the $L$–$L'$ direction with LV polarization. The bulk conduction band C is clearly resolved, while the SRS appears with reduced intensity. (d) Schematic band reconstruction. (e–f) Fermi surfaces at different photon energies distinguish surface (circular) from bulk (elongated) states. (g-h) CDW folding effects as elliptical bulk pockets and zone center replicas. (i) Schematic Fermi surface illustrating bulk ellipsoids and surface circular dispersion. Dashed green line in (g) marks the analyzer slit direction. The Fermi surface contours are obtained by integrating the spectral weight within a 10 meV window around the Fermi level. ARPES measurements are conducted at 50 K sample temperature. The color scale represents the photoemission intensity (arb. units) and is common to all panels.}
\end{figure*}

Here, we demonstrate the emergence of a SRS in 1T-TiSe$_2$. We identify this state as an SRS, localized at the surface but energetically embedded within the bulk continuum, distinguishing it from a topological surface state or a simple CDW-backfolded replica. The SRS shows behavior consistent with quasi-two-dimensional character, and disappears around 160~K. Notably, while bulk band structure calculations for TiSe$_2$ are abundant, surface-resolved electronic structure calculations have not been previously  reported. By combining $\mu$-ARPES, photon energy–dependent mapping, and first principles DFT+$U$ slab calculations, we suggest that the SRS arises from surface selective interaction between folded valence and conduction bands, tuned by correlation effects. The essential ingredients are band folding from symmetry breaking and correlation enhanced band structure tuning which are not unique to TiSe$_2$. Therefore, similar states could occur in other layered CDW systems such as 1T-TaS$_2$~\cite{ritschel2015orbital,wang2024dualistic}, 2H-NbSe$_2$~\cite{kiss2007charge,borisenko2009two}, and rare-earth tritellurides~\cite{brouet2008angle,moore2010fermi}. Our work therefore positions 1T-TiSe$_2$ not only as a singular puzzle case but also as a possible prototype for a broader class of correlation driven surface metallic states in symmetry broken layered materials.

\begin{figure*}[t]
	\centering
	\includegraphics[width=0.95\textwidth]{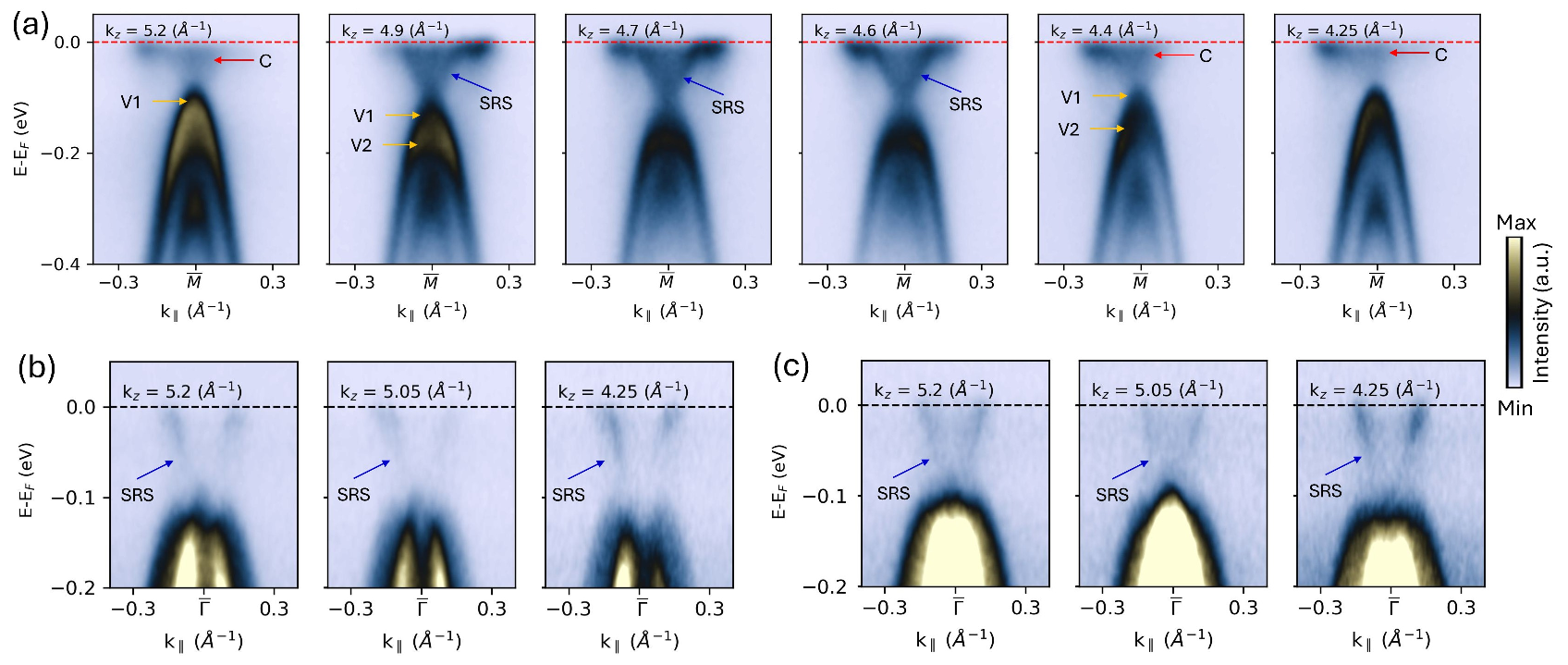} 
	
	\caption{
		(a) ARPES spectra measured at the $\overline{M}$ point across multiple photon energies, corresponding to different $k_z$ values. The spectral weight of the SRS exhibits strong photon energy dependence, becoming weak at certain $k_z$ slices, while the C band becomes more prominent (highlighted by red arrows).
		(b, c) Photon energy–dependent ARPES measurements at the $\overline{\Gamma}$ point, where the SRS remains observable. ARPES data in (b) and (c) are recorded along the $\overline{M}$--$\overline{\Gamma}$--$\overline{M}$' and $\overline{K}$--$\overline{\Gamma}$--$\overline{K}$', respectively. All ARPES measurements, here, were performed at a sample temperature of 50 K. The color scales represent photoemission intensity (arb. units).}
	
\end{figure*}

\section*{ARPES Evidence for Surface–Bulk Separation in the CDW State}

Previous ARPES studies of 1T-TiSe$_2$ identify a V-shaped band at the $M$-point and assigned it to a bulk Ti d$_{z^{2}}$ orbital with strong k$_z$ dispersion~\cite{watson2019orbital}. However, the origin of this feature has remained ambiguous due to the superposition of bulk and surface contributions that are challenging to separate. Using high-resolution $\mu$-ARPES with systematic polarization and photon-energy control, we resolved two distinct bands near the Fermi level as (i) a broad, k$_z$-dispersive bulk conduction band (C), and (ii) a sharp V-shaped SRS that does not show a clear k$_z$-dependent energy shift. Therefore, in Fig.~1, we first identify a sharp V-shaped SRS, spanning the bulk band gap using polarization and geometry dependent ARPES at a sample temperature of 50~K. Measurements were first taken at 119~eV, corresponding to the bulk $k_z = A$ plane (Fig.~1(a–c)), ensuring all spectra were recorded from the same $k_z$-point to maximize matrix element sensitivity to different orbital characters.

Polarization-dependent measurements reveal distinct orbital contributions near the $L$-point. The SRS appears under both linear horizontal (LH) and vertical (LV) light, with LH preferentially enhancing the folded valence band V2 (Fig.~1(a)) and LV highlighting V1 (Fig.~1(b)). Under LH, weak features appear alongside the SRS (dashed white parabolas in Fig.~1(a)), suggesting localized hybridization with the conduction band C. Extending the measurement along $L$--$L'$, LV polarization also resolves a shallow parabolic like C band which appears enhanced along this particular high symmetry direction (Fig.~1(c)). Energy distribution curves (EDCs) further illustrate the separation between the SRS and the bulk conduction band, with geometry-dependent spectral weight highlighting their distinct nature (see Fig.~S1). Together, these observations are consolidated in the schematic of Fig.~1(d).

To further distinguish between surface and bulk contributions, we examine photon-energy- and geometry-dependent Fermi surface maps. At a photon energy of 119~eV with LH polarization, circular Fermi pockets centered at the bulk $L$ points are observed, accompanied by faint, elongated intensity features extending along the $L$--$A$ direction (Fig.~1(e)), consistent with bulk-like dispersions. At 99 eV photon energy, the Fermi surface near the surface-projected $\overline{M}$ point becomes more circular, consistent with an increased contribution from the surface-resonant state, while the elongated features associated with the bulk conduction band are suppressed (Fig.~1(f)), signifying that the corresponding states predominantly originate from the surface.. Aligning the analyzer slit along $\overline{M}$–$\overline{M}'$ suppresses surface states at the Fermi level, leaving bulk C band ellipsoids dominant (Fig.~1(g)). This supports a composite Fermi surface at $\overline{M}$ as combination of bulk derived ellipsoids plus surface induced circular pockets. Furthermore, CDW folding of these features forms the six $\overline{M}$ centered ellipsoids at the $\overline{\Gamma}$, producing mirror dispersions (Fig.~1(h)).

These polarization, geometry, and photon energy–dependent results indicate the  orbital selective, surface localized nature of the SRS and reveal a clear surface bulk separation in the CDW phase of 1T-TiSe$_{2}$ as also illustrated in (Fig.~1(i)). The apparent flower like  contour at $\overline{\Gamma}$ arises because all three inequivalent $\overline{M}$ centered conduction ellipsoids are folded onto the zone center and overlap, whereas at a given $\overline{M}$ point only 
the primary valley remains bright; the symmetry equivalent replicas fold to other $\overline{M}$ points but their ARPES weight is likely suppressed due to the matrix element effect. This reasoning also supports that the sharp V-shaped band observed in addition to the bulk C band is unlikely to be explained as a simple backfolded replica. Hence, the data indicate how symmetry breaking in layered materials can lead to formation of emergent low energy excitation.

\begin{figure*}[t]
	\centering
	\includegraphics[width=0.65\textwidth]{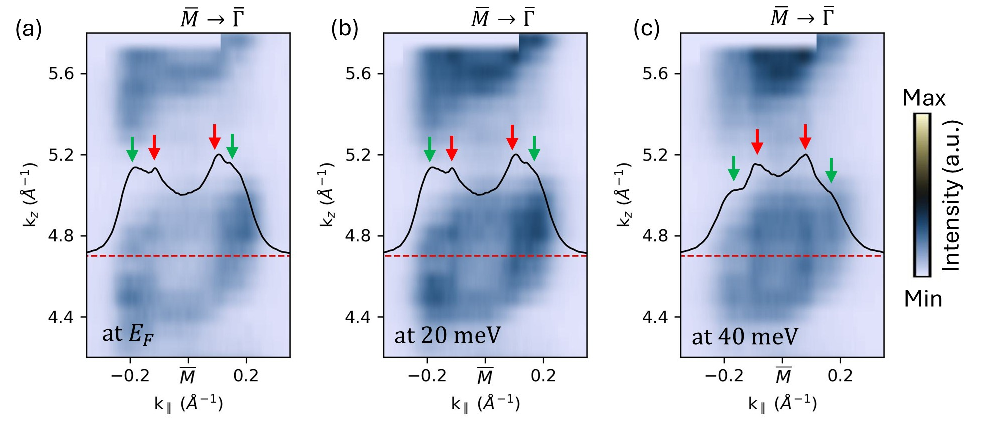} 
	\caption{
		(a–c) $k_z$ vs $k_\parallel$ ARPES constant-energy maps at the $\overline{M}$ point, taken at (a) $E_F$, (b) $-20$ meV, and (c) $-40$ meV binding energies, each integrated within a 10 meV window. Momentum distribution curves (MDCs) extracted along the dashed lines are overlaid. At all three energies, the MDCs reveal two distinct features: a sharp peak associated with the SRS (red arrows) and a broader peak corresponding to the bulk conduction band, C (green arrows).}
	
\end{figure*}

\section*{Photon Energy Dependence Reveals Two Dimensional Surface Resonance}

To assess the dimensional character of the SRS and distinguish it from bulk states, we performed photon-energy–dependent ARPES at the $\overline{M}$-point (Fig.~2(a)). Photon‑energy variation in ARPES is a well‑established method for probing k$_z$ dispersion and orbital selectivity~\cite{damascelli2003angle,strocov2003intrinsic}. However, in the present case, the SRS strongly overlaps with the bulk conduction band and their energy separation is small, making it difficult to extract precise peak positions as a function of photon energy. The measured feature exhibits pronounced spectral-weight modulation with photon energy, yet SRS does not show a clear or systematic k$_z$ dispersion within experimental resolution. Unlike isolated topological or trivial surface states, which reside entirely outside the projected bulk band manifold~\cite{hasan2010colloquium}, the observed state partially overlaps in energy and momentum with bulk-derived bands, particularly the C and V1 bands.

Such coexistence with bulk dispersions is the defining property of a surface resonant state which is a surface-localized state embedded within the bulk continuum, with finite hybridization to bulk states~\cite{kneedler1990surface,ehlen2018direct}. Within this framework, a genuine surface resonance is characterized by (i) surface localization, (ii) no clear k$_z$-dependent energy shift, and (iii) energetic embedding within the bulk continuum—criteria that are simultaneously satisfied by the observed feature. The energetic overlap, combined with photoemission matrix-element effects, leads to strong photon-energy-dependent visibility and complicates its experimental isolation. These characteristics distinguish the feature from both a purely two-dimensional surface state and a CDW-backfolded bulk replica. We therefore classify it as a genuine SRS. Its partial embedding in the bulk continuum likely explains why its surface-localized character was not fully recognized in earlier ARPES studies focused primarily on the bulk electronic structure~\cite{cercellier2007evidence,monney2009temperature,bovet2004pseudogapped,nohara1991angle,vydrova2015three}.

To disentangle these effects, we carried out systematic scans across a broad range of photon energies. These measurements reveal a pronounced photon-energy-dependent intensity modulation. The SRS displays strongly modulated spectral weight, becoming faint at certain momenta (e.g., $k_z = 5.2$~\AA$^{-1}$ and $4.25$~\AA$^{-1}$) where bulk features dominate, but is clearly resolved at others (blue arrows, Fig.~2(a)). LV polarization modifies the relative spectral weight of the SRS and bulk states through matrix-element effects, leading to photon-energy- and geometry-dependent contrast (figs.~S2–S3). As shown in Fig. S3, the SRS persists across all photon energies, while its visibility varies due to matrix-element effects, resulting in changes in contrast rather than a pronounced enhancement at a specific photon energy. These complementary approaches consistently confirm the SRS across the probed $k_z$ range, providing a robust basis for examining its behavior in other regions of the Brillouin zone.

Due to the PLD, the $\overline{\Gamma}$-centered valence bands fold onto $\overline{M}$, while the $\overline{M}$-centered conduction band folds onto $\overline{\Gamma}$. Consequently, the SRS is also expected to appear at the zone center, providing a means to identify it at the same $k_z$ as the $\overline{M}$ point. Although weaker in this momentum region compared to the valence states, the SRS remains detectable along both $\overline{M}$–$\overline{\Gamma}$–$\overline{M}'$ and $\overline{K}$–$\overline{\Gamma}$–$\overline{K}'$ directions (Fig.~2(b, c)). The resulting $k_z$–$k_\parallel$ maps do not show a clear or systematic $k_z$ dispersion, consistent with a two-dimensional character of the SRS near the zone center (Figs.~S4–S5).

However, a direct assessment of $k_z$ dispersion based solely on peak positions is not straightforward in the present case. While photon-energy-dependent measurements are often used to probe $k_z$ dispersion, the strong overlap between the SRS and bulk states and their limited energy separation prevent a reliable extraction of peak positions. The observed photon-energy--dependent intensity modulation is therefore attributed to matrix-element and final-state effects rather than intrinsic dispersion. Accordingly, the identification of the SRS as a quasi-two-dimensional state is based on a combination of complementary measurements, including its polarization dependence, momentum-space geometry, and behavior at the $\bar{\Gamma}$ point, together with slab DFT+$U$ calculations.

Further evidence for the distinct nature of the SRS and bulk conduction band, C, is provided by constant-energy analysis at the $\overline{M}$ point (Fig.~3(a-c)), where momentum distribution curves (MDCs) reveal two coexisting features at the same momentum: a sharp peak associated with the SRS and a broader peak corresponding to the bulk conduction band. The clear difference in their peak profiles provides direct evidence for two separate electronic states. The coexistence of bulk and surface contributions at the same momentum leads to spectral overlap between these features, thereby providing a likely explanation for why earlier work~\cite{watson2019orbital} attributed the entire V-shaped feature to a bulk band.

\begin{figure*}[t]
	\centering
	\includegraphics[width=0.95\textwidth]{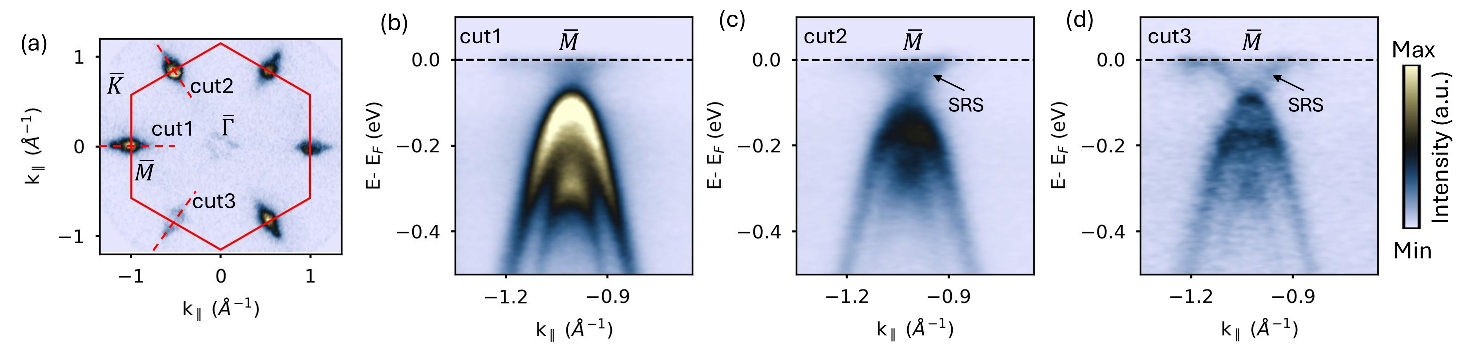} 
	
	\caption{
		(a) Fermi surface map of 1T-TiSe$_2$ measured at 50 K using 99 eV LH-polarized photons, showing the hexagonal Brillouin zone and three representative momentum cuts (cut1–cut3, red dashed lines). (b–d) ARPES intensity plots along cut1, cut2, and cut3, respectively. The surface resonant state (SRS) is clearly resolved along certain momentum cuts (cut2 and cut3) but strongly suppressed along others (cut1), despite identical photon energy conditions. This behavior demonstrates that the visibility of the SRS is governed by matrix element effects.}
	
\end{figure*}

\section*{Momentum-dependent spectral weight of the surface resonant state}

To further elucidate the origin of the SRS, we examine its dependence on measurement geometry at fixed photon energy. Fig.~4(a) shows the Fermi surface map of 1T-TiSe$_2$ measured at 50~K using 99~eV LH polarized photons, with three representative momentum cuts (cut1--cut3) indicated. Fig.~4(b, d) present the corresponding ARPES intensity plots along these cuts.

Despite identical photon energy and polarization conditions, the SRS exhibits a strong momentum dependence. It is clearly resolved as a sharp, dispersive feature particularly along cut3, while it is strongly suppressed along cut1. This variation occurs even though all cuts correspond to the same k$_z$, indicating that the SRS is not governed solely by photon energy or $k_z$.

This behavior is naturally explained by photoemission matrix element effects, which modulate the spectral weight depending on the orbital character of the states and the experimental geometry. As a result, the SRS can be selectively enhanced or suppressed depending on the measurement configuration. These results show that the absence or weak intensity of the SRS in specific cuts does not imply its intrinsic absence, but rather reflects matrix-element-driven modulation of its spectral weight.

Our results are in fact consistent with the earlier reports~\cite{watson2019orbital} where the circular $k_z$ modulation we observe at the Fermi level (Fig.~3) reproduces the reported bulk dispersion. However, by extending the measurement to higher binding energy and extracting momentum MDCs, we resolve an additional sharp peak that exhibits no apparent $k_z$ dispersion and gains spectral weight as the bulk band fades. Our data thus confirm the established bulk electronic structure while revealing a previously unnoticed surface resonance.

Having established that the SRS is spectroscopically distinct from bulk bands, we next establish the intrinsic nature of the SRS by examining multiple cleaves and performed spatially resolved $\mu$-ARPES (Figs.~S6–S7). The SRS was consistently observed irrespective of surface location, cleaving conditions, or sample history, thereby ruling out extrinsic origins such as surface doping, interlayer decoupling, or electronic inhomogeneity.

Surface quality is further confirmed by scanning tunneling microscopy (STM) and core-level spectroscopy, which definitively establish the pristine 1T termination of the cleaved surface. Atomic-resolution topography (Figs.~S8(a)) resolves a hexagonal lattice with spacing $a = 0.352$~nm, matching the bulk 1T phase. The commensurate $2\times2$ CDW modulation yields a real-space period of $\lambda = 0.713$~nm. Fast Fourier transform (Figs.~S8(b)) reveals Bragg peaks at $\mathbf{q}_\mathrm{lattice} = 3.280$~nm$^{-1}$ and CDW satellites at $\mathbf{q}_\mathrm{CDW} = 1.623$~nm$^{-1}$, giving a ratio $|\mathbf{q}_\mathrm{lattice}|/|\mathbf{q}_\mathrm{CDW}| = 2.02$, confirming strict $2\times2$ commensurability. Complementary Se $3d$ core-level spectroscopy (Figs.~S8(c)) exhibits a single, sharp spin-orbit doublet with no additional components or asymmetric broadening. The absence of extra peaks, commonly associated with Se deficiency, surface reconstructions, or poly-type inclusions, confirms stoichiometric composition and a pristine 1T surface. Together, these observations definitively identify the surface as bulk-terminated 1T-TiSe$_2$, ruling out alternative terminations or non-stoichiometry as the origin of the SRS. These results confirm the structural and chemical integrity of the cleaved surface, eliminating extrinsic artifacts as an explanation for the observed SRS.

\begin{figure*}[t]
	\centering
	\includegraphics[width=0.95\textwidth]{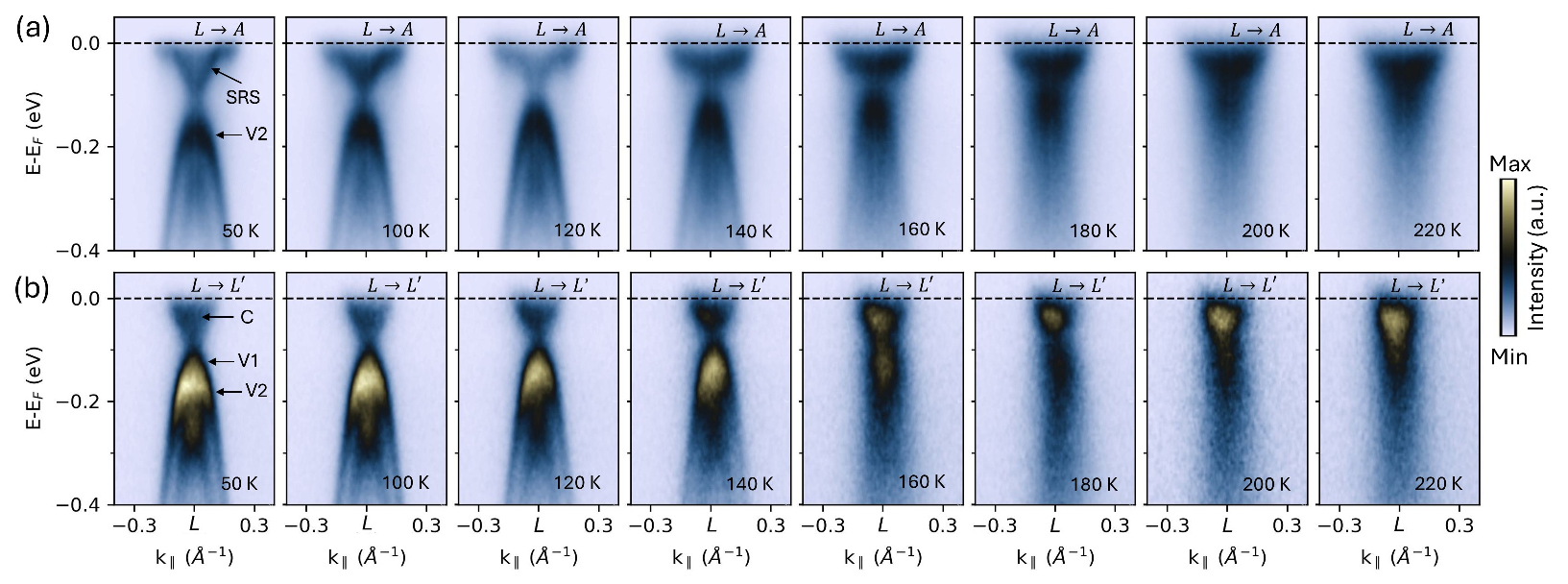} 
	
	\caption{
		(a, b), Temperature-dependent ARPES spectra along the $L$--$A$ and $L$--$L$' directions, showing the progressive suppression of the SRS with increasing temperature. The sharp V-shaped dispersion vanishes between 140~K and 160~K while the broad conduction band becomes increasingly dominant toward higher temperatures, indicating the collapse of the correlation driven surface resonance well below the bulk CDW transition at $T_{CDW} \approx 202$~K. The color scale represents the photoemission intensity (arb. units) and is common to all panels.}

\end{figure*}

\section*{Temperature Induced Suppression of Surface Resonance}

To track the evolution of the low-energy electronic structure across the CDW transition, we performed temperature-dependent ARPES measurements along the $L$--$A$ and $L$--$L'$ directions, where the SRS and conduction (C) band are best resolved, respectively (Fig.~5(a, b)). As temperature increases, the sharp V-shaped SRS gradually weakens and vanishes between 140~K and 160~K. Its disappearance coincides with a progressive transfer of spectral weight to the broad, bulk-like C band, which becomes increasingly prominent at higher temperatures (Fig.~5(a, b)).

This evolution can be understood as a redistribution of spectral weight between the bulk conduction band C and the SRS. At low temperature, the SRS dominates low energy spectra, producing the sharp V‑shaped dispersion. Importantly, our photon-energy- and polarization dependent measurements directly reveal that the SRS and the bulk C band coexist at the same momentum, with both features simultaneously visible under favorable matrix-element conditions. The suppression of the SRS near 160~K is independently confirmed by high-resolution laser ARPES, which isolates the resonance from bulk contributions (fig.~S9). This data clearly reveals the transition from sharp SRS as the dominant low energy feature to broader bulk conduction band at higher temperatures. As temperature increases and the SRS collapses, spectral weight transfers back to the broad, weakly dispersive bulk C band. This switching of spectral weight naturally explains the apparent evolution from a sharp V‑shaped feature at low temperature to a broad, relatively flat conduction band at higher temperature. In previous ARPES studies, this change was implicitly attributed to a temperature dependent modification of the conduction band itself because the surface contribution was not recognized~\cite{watson2019orbital,ou2024incoherence}. Our results show instead that the underlying bulk C band remains essentially unchanged. Without such a spectral-weight-transfer mechanism, it would be unlikely to justify a strong renormalization of the normal phase conduction band dispersion in to V-shaped one solely by temperature.

 In parallel with the evolution of the SRS and the conduction band, the folded valence band V2 exhibits a distinct temperature dependence. As temperature increases, V2 shifts toward the Fermi level, with the magnitude of this shift saturating near 160~K as is also evident in the EDCs presented in Fig.~S10. Above this temperature, the spectral weight of V2 progressively diminishes, reflecting the gradual weakening of CDW  induced band folding. This behavior mirrors the temperature scale associated with the SRS collapse, indicating that both the valence band reconstruction and the surface resonance are governed by the same underlying evolution of CDW driven band structure modification. The simultaneous suppression of V2 and the SRS therefore provides a consistent picture in which the near-degeneracy between folded valence and conduction states, essential for sustaining the surface resonance, is lost as the system approaches the incoherent regime above $\approx$ 160~K.

\begin{figure*}[t]
	\centering
	\includegraphics[width=0.95\textwidth]{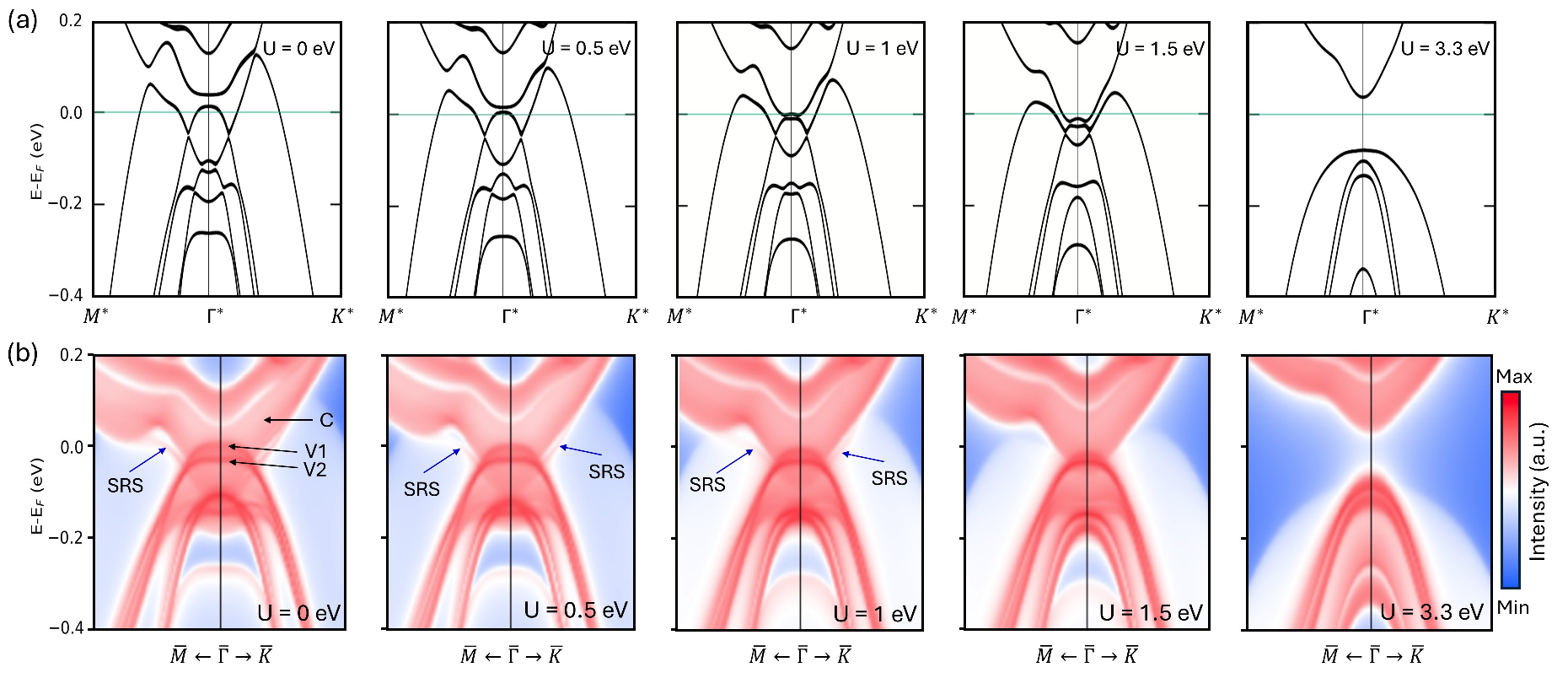} 
	
	\caption{
		(a) Bulk DFT+$U$ band structures of the distorted CDW phase, showing the evolution of the folded Se $p$-derived valence bands and Ti $d$-derived conduction band as a function of on-site Hubbard interaction $U$ applied to Ti $d$ orbitals. For moderate $U$, these bands become nearly degenerate near the zone center, facilitating orbital hybridization without opening a full gap. At larger $U$, however, a full insulating gap emerges. (b) Slab DFT+$U$ calculations of the same CDW phase reveal a SRS near the Fermi level at low $U$, overlapping in both energy and momentum with the folded bulk bands near $\overline{\Gamma}$. This state fades at larger $U$ as hybridization between conduction (C) and valence bands (V1/V2) diminishes.}
	
\end{figure*}

The temperature‑dependent evolution of the spectral features observed here, including changes in the apparent separation between the conduction and folded valence bands, resembles trends reported in earlier ARPES studies of 1T-TiSe$_2$~\cite{monney2009temperature,kidd2002electron,ou2024incoherence,chen2016hidden}. Fig.~S10 shows temperature-dependent spectra acquired at the $\overline{M}$ point under an ARPES geometry in which the bulk conduction band dominates the spectral weight while the SRS remains faint. This configuration highlights how photon‑energy, polarization, and k$_z$ resolved measurements are essential for isolating the surface resonance from overlapping bulk states. Without such selectivity, the SRS can be easily obscured, leading earlier studies to interpret the temperature evolution solely in terms of changes to the bulk conduction band.

Furthermore, a long-standing puzzle in 1T-TiSe$_2$ is the transport anomaly near 160~K~\cite{rossnagel2002charge,miyahara1996tunnelling,craven1978mechanisms}. This temperature scale is notable because it lies well below the bulk CDW transition (T$_{CDW}$ $\approx$ 202~K), where ARPES confirms that band folding persists, yet coincides with the collapse of the SRS we observe. Our identification of a surface resonance that vanishes at this same temperature provides a crucial spectroscopic link to this anomaly. While transport in 1T-TiSe$_2$ is dominated by bulk carriers, precluding a direct surface-state conduction path, the SRS can be understood as a sensitive probe of the low-energy electronic coherence in the material. Its disappearance at 160~K signals a fundamental change in the electronic scattering and coherence that is not captured by the persistence of the CDW order alone. Consistent with recent reports of a coherent-to-incoherent crossover near this temperature~\cite{ou2024incoherence}, we interpret the SRS collapse as a direct manifestation of this bulk electronic decoherence. The transport anomaly, therefore, is not caused by the SRS itself, but both phenomena share a common origin in a temperature-driven evolution of the low‑energy band structure and electronic coherence.

This interpretation naturally connects to the broader temperature-dependent evolution of the CDW state and its associated electronic structure. The suppression of the SRS with increasing temperature occurs in a temperature range where the CDW order parameter has been reported to weaken in prior studies. Recent theoretical and experimental work indicate that the electronic structure of TiSe$_2$ is governed by CDW-induced band hybridization and electron–phonon coupling, rather than purely excitonic effects~\cite{pashov2025tise2,fragkos2026electron}. Furthermore, the development of CDW order is known to be gradual, with electronic coherence and amplitude modes becoming more pronounced at lower than $T_{CDW} \approx 202$~K temperatures~\cite{duong2017raman,cui2017raman,wang2018large,tang2022growth}. In this context, although the PLD sets in at $T_{{CDW}}$, the associated electronic coherence and long-range CDW order develop progressively upon cooling. The temperature-dependent evolution of the SRS therefore reflects the weakening of CDW-induced hybridization as electronic coherence is reduced with increasing temperature.

\section*{Correlation Tuned Mechanism for Surface Resonances in Layered CDW Systems}

Our ARPES measurements reveal that the disappearance of the SRS at elevated temperatures coincides with reduced energetic overlap between the folded valence and conduction bands. To determine whether this reflects a material specific peculiarity or a more general phenomenon, we performed DFT+$U$ calculations for both bulk and slab geometries of the CDW distorted 1T-TiSe$_{2}$.

In the bulk calculations (Fig.~6(a)), increasing the on site Hubbard interaction $U$ on Ti $d$ orbitals progressively enlarges the separation between the folded valence bands (V1, V2) and the conduction band (C), consistent with correlation enhanced band repulsion. At large $U$, this separation is sufficient to open a full bulk gap, whereas at moderate $U$ the conduction and valence band edges approach near degeneracy at the Brillouin zone center, creating favorable conditions for strong interband interaction. This near degeneracy in the bulk electronic structure is a key condition for the emergence of the surface state. 

However, while the on-site Hubbard $U$ controls the overall energy separation between conduction and valence bands in a momentum-independent manner, the manifestation of the resulting hybridization differs between the $\Gamma$ and $M/L$ points due to the underlying band structure and CDW folding conditions. At the $M/L$ points, the conduction band minimum has strong Ti $d$ character and directly hybridizes with the folded valence bands, leading to a more pronounced SRS. In contrast, at $\Gamma$, the relevant states arise primarily through CDW-induced backfolding of bands from $M$, and the hybridization occurs more indirectly, resulting in a comparatively weaker spectral weight.

Slab calculations (Fig.~6(b)) show that near degeneracy fosters a surface localized resonance near the Fermi level, overlapping in both momentum and energy with folded bulk states. This resonance disappears once the separation exceeds a correlation dependent threshold. The $Z_2$ invariant is also found to be trivial, and spin resolved ARPES shows no net spin polarization (fig.~S11), excluding topological origins and pointing instead to a correlation driven, surface selective reconstruction enabled by CDW folding.

The surface presents a distinct electronic environment compared to the bulk, characterized by reduced dimensionality, modified boundary conditions, and reconstruction. Rather than assuming a simple monotonic change in the bare Hubbard interaction, these factors effectively renormalize the relative alignment between folded valence and conduction bands at the surface. Within the DFT+$U$ framework, the parameter $U$ primarily controls correlation-driven band separation rather than directly representing the microscopic screened Coulomb interaction. Our slab calculations therefore demonstrate that the emergence of a surface-localized resonance is governed by correlation tuned electronic structure, such that when the folded states approach near degeneracy, a surface resonance appears, whereas increased separation suppresses it. This interpretation emphasizes the sensitivity of surface electronic structure to correlation-induced band renormalization without requiring a strict quantitative modification of the bare $U$ at the surface.

\section*{Discussion}

The formation of a surface resonance in 1T-TiSe$_2$ below ~160~K, well separated from the bulk CDW transition, parallels the behavior of correlated surface states in other layered quantum materials. In Sr$_2$RuO$_4$, for example, a $\sqrt{2}\times\sqrt{2}$ surface reconstruction produces a surface-derived band with a strongly enhanced effect mass, distinct from the bulk electronic structure~\cite{damascelli2000fermi,tamai2019high}. In 1T-TiSe$_2$ the $2\times2$ CDW reconstruction plays an analogous role. At the surface, CDW-induced folding brings the Se $p$ and Ti $d$ states into near degeneracy, and in this regime the modified surface electronic environment promotes a redistribution of spectral weight from bulk like bands into a coherent surface resonance. These observations illustrate how the interplay of symmetry-breaking reconstructions and surface-specific electronic renormalization can generate emergent metallic channels in otherwise gapped layered materials, offering a general framework for engineering correlated surface states in van der Waals quantum systems.

1T-TiSe$_2$, the SRS arises from two essential ingredients: (i) the CDW distortion folds Ti $d$-derived conduction and Se $p$-derived valence states into the same momentum sector, and (ii) their small energy separation allows correlations or surface effects to tune them into energetic proximity. The surface-localized nature of the SRS is supported by the lack of a clear k$_z$ dispersion, its reproducibility across photon energies and polarizations, and its distinct temperature scale. Notably, the SRS appears only well below the bulk CDW transition temperature $T_{CDW} \approx 202$~K, excluding a simple backfolded-band interpretation. Instead, it reflects a selective reconstruction confined to the surface, where reduced dimensionality and surface-specific renormalization stabilize a correlated metallic channel.

More broadly, 1T-TiSe$_2$ provides a model system for a general mechanism of correlation‑tuned surface‑resonance formation. Layered CDW materials such as 1T-TaS$_2$, 2H-NbSe$_2$, rare-earth tritellurides, and kagome metals (AV$_3$Sb$_5$) share the same essential ingredients-zone folding, moderate correlations, and surface‑modified electronic structure, making them natural candidates for analogous surface resonances. Such states could mediate coupling between charge order, superconductivity, and topology, or serve as tunable metallic channels in otherwise gapped systems.

Our results extend the established bulk picture of the CDW electronic structure in 1T-TiSe$_2$ by explicitly incorporating the surface degree of freedom. While the underlying bands are bulk‑derived and consistent with previous reports, slab calculations reveal that CDW-induced folding can stabilize a surface resonance when conduction and valence states approach energetic overlap at the surface. This finding highlights that surface‑localized resonances can naturally arise in layered CDW materials without invoking topological or extrinsic mechanisms, underscoring the importance of treating surface and bulk electronic structure on equal footing.

By identifying the minimal symmetry and energetic conditions required for surface‑resonance formation, this work establishes a general framework for realizing and controlling correlated surface states in broken‑symmetry quantum materials. It shows that even in well‑studied systems, the surface can host qualitatively new phases, transforming a bulk CDW semiconductor into a system supporting a correlated surface resonance.

\section*{Methods}

\subsection*{Sample sourcing and preparation}

Single crystals of 1T-TiSe\textsubscript{2} were obtained commercially from 2Dsemiconductors.

\subsection*{ARPES experiments}

$\mu$-ARPES and spin-resolved ARPES experiments were performed at the ESM beamline of NSLS-II (21-ID-1)~\cite{rajapitamahuni2024electron} and the APE-LE beamline of Elettra, respectively, using DA30 Scienta electron spectrometers at both facilities. Samples were cleaved \textit{in situ}, and the base pressure in the photoemission chambers was maintained below $3 \times 10^{-11}$~Torr.

The photon beam spot size was approximately $5~\mu\mathrm{m}^2$ at 21-ID-1 and $150 \times 50~\mu\mathrm{m}^2$ at APE-LE. Synchrotron radiation was incident at an angle of $55^\circ$ at 21-ID-1. LV polarization was oriented parallel to both the sample surface and the analyzer slit, while LH polarization lay in the plane of incidence, enabling selective probing of different orbital symmetry components.

The laser ARPES experiments were carried out with hv = 6.3 eV and R4000 Scienta electron spectrometer at the Research Institute for Synchrotron Radiation Science (HiSOR), Hiroshima University. The expected energy resolution was approximately 5 meV with ~10 $\mu$m incident photon beam spot diameter \cite{iwasawa2017development}.

The out-of-plane momentum component $k_z$ was estimated using the free-electron final-state approximation~\cite{damascelli2003angle}
\[
k_z = \sqrt{\frac{2m}{\hbar^2}\left(E_{\mathrm{kin}}\cos^2\theta + V_0\right)},
\]
where $E_{\mathrm{kin}}$ is the photoelectron kinetic energy, $\theta$ is the emission angle relative to the surface normal, $V_0$ is the inner potential (taken to be 13.5~eV for 1T-TiSe$_2$~\cite{watson2019orbital}), $m$ is the free electron mass, and $\hbar$ is the reduced Planck constant.

\section{DFT Calculations}

Plane-wave DFT calculations were carried out using the Quantum ESPRESSO package~\cite{giannozzi2009quantum,van2018pseudodojo}. We employed the relativistic norm-conserving pseudopotential taken from PseudoDojo~\cite{van2018pseudodojo,hamann2013optimized} to account for spin-orbit coupling calculation, and the Perdew-Burke-Ernzerhof generalized gradient approximation (PBE-GGA)~\cite{perdew1996generalized} as the exchange-correlation functional. The Brillouin zone of the distorted phase was sampled with a 4$\times$4$\times$4 Monkhorst-Pack grid in an enlarged 2$\times$2$\times$2 supercell. A plane-wave kinetic energy cutoff of 80 Ry and a Gaussian smearing width of 0.01 Ry were used to ensure smooth integration along the Fermi surface. Subsequently, to calculate the surface states, we first performed wannierization using the Wannier90 package~\cite{marzari1997maximally}. This step yielded a maximally localized tight-binding Hamiltonian, which was then used as input for the surface state calculations carried out with WannierTools.

\section{Scanning tunneling microscopy}

Single-crystal 1T-TiSe$_2$ was cleaved at room temperature in an ultrahigh-vacuum chamber ($\sim$$2\times10^{-10}$ Torr) and immediately transferred to the STM stage, where it was cooled to 10.5 K. Topographic and spectroscopic data were acquired with PtIr tips and reproduced across multiple cleaves and tips to confirm intrinsic behavior.

Topographs obtained in constant-current mode revealed large, atomically flat terraces with randomly distributed native defects. A triangular lattice with nearest-neighbor spacing of 0.352 nm was resolved, consistent with the bulk lattice constant, along with a commensurate $2\times2$ CDW modulation.

Fast Fourier transforms (FFTs) of the images, analyzed with \textit{WSxM} software~\cite{horcas2007wsxm}, confirm the commensurate CDW by comparing the CDW modulation vectors $\mathbf q_\mathrm{CDW}$ with the atomic Bragg vectors $\mathbf q_\mathrm{lattice}$. The ratio $|\mathbf q_\mathrm{lattice}|/|\mathbf q_\mathrm{CDW}|$ extracted from peak positions in the FFT was 2.

\section*{Acknowledgments}

This research used resources at the ESM (21-ID-1) beamline of the National Synchrotron Light Source II, a U.S. Department of Energy (DOE) Office of Science User Facility operated by Brookhaven National Laboratory under Contract No.~DE-SC0012704. This work was also supported by a research grant from Xiamen University Malaysia (Grant No.~IORI/0007). P.M.S., P.M., J.F., and I.V. acknowledge support from the EUROFEL-ROADMAP ESFRI project funded by the Italian Ministry of Education, University, and Research. Part of this work was carried out within the framework of the Nanoscience Foundry and Fine Analysis (NFFA-MUR Italy Progetti Internazionali) facility. The laser ARPES measurements were performed with the approval of the Proposal Assessing Committee of HiSOR (Proposal No. 23AU013).

\section*{Author Contributions}

T.Y. conceived and supervised the research. T.Y., A.R., A.K., B.S., and E.V. conducted the ARPES experiments at the 21-ID-1 beamline. T.Y, S.I., and K.S performed the laser based ARPES experiments at HISOR. P.M.S., P.M., J.F., and I.V. carried out the spin-resolved ARPES experiments at the APE-LE beamline. M.A.F. and S.M.H conducted the STM data. Y.S.N. performed the theoretical calculations under the supervision of H.Q.W and J.C.Z. T.Y. wrote the manuscript with input from all authors. All authors contributed to the scientific discussion and interpretation of the results.

\section*{Data availability}
The data that support the findings of this study are available from the corresponding author upon request.

\section*{Competing Interests}
The authors declare no competing interests.

\bibliography{references}

\end{document}